# Picosecond expansion in LaAlO$_3$ resonantly driven by infrared-active phonons

Jakob Gollwitzer,[1,*] Jeffrey Z. Kaaret,[2,*] Y. Eren Suyolcu,[1] Guru Khalsa,[1,3] Rylan C. Fernandes,[1] Oleg Gorobtsov,[1] Sören Buchenau,[2] ChanJu You,[4] Jayanti Higgins,[2] Ryan S. Russell,[5] Ziming Shao[1], Yorick A. Birkhölzer[1], Takahiro Sato,[6] Matthieu Chollet,[6] Giacomo Coslovich,[6] Mario Brützam,[7] Christo Guguschev,[7] John W. Harter,[5] Ankit S. Disa,[2] Darrell G. Schlom,[1,7,8] Nicole A. Benedek,[1,†] and Andrej Singer[1,‡]

[1]Department of Materials Science and Engineering, Cornell University, Ithaca, New York 14853, USA2

[2]School of Applied and Engineering Physics, Cornell University, Ithaca, New York 14853, USA

[3]Department of Physics, University of North Texas, Denton, Texas 76203, USA

[4]Department of Physics, Cornell University, Ithaca, NY 14853, USA

[5]Materials Department, University of California, Santa Barbara, California 93106, USA

[6]The Linac Coherent Light Source, SLAC National Accelerator Laboratory, Menlo Park, CA 94025 USA

[7]Leibniz-Institut für Kristallzüchtung, Max-Born-Straße 2, 12489 Berlin, Germany

[8]Kavli Institute at Cornell for Nanoscale Science, Cornell University, Ithaca, NY 14853, USA

*These authors contributed equally to this work.

†nbenedek@cornell.edu, ‡asinger@cornell.edu

## Abstract

We investigate the ultrafast structural dynamics of LaAlO$_3$ thin films driven by short mid-infrared laser pulses at 20 THz. Time-resolved X-ray diffraction reveals an immediate lattice expansion and an acoustic breathing mode of the film. First-principles theory and a spring-mass model identify the direct coupling between coherently driven infrared-active phonons and strain as the underlying mechanism. Time-resolved optical birefringence measurements confirm that the amplitude of this acoustic mode scales linearly with the pump fluence, which agrees with the theory. Furthermore, time-resolved X-ray diffuse scattering indicates that THz excitation enhances crystallinity by inducing a non-thermal increase in structural symmetry originating from preexisting defects. These findings highlight the potential of a multimodal approach—combining time-resolved X-ray and optical measurements and first-principles theory—to elucidate and control structural dynamics in nanoscale materials.

*Introduction*–The advent of intense laser sources in the terahertz (THz) frequency range has enabled the exploration of non-equilibrium phenomena in quantum materials by coherently driving vibrational phonon modes. Researchers can now control functional properties such as superconductivity[1-3], ferroelectricity[4,5], and magnetism[6-8] on picosecond time scales. $LaAlO_3$ is a well-known perovskite compound that serves as the host for a variety of interesting phenomena in bulk[9] and interfaces[10]. Recently, two experiments employed a THz-pump resonant with an infrared-active phonon and an optical probe to observe ultrafast changes in reflectivity, detecting oscillations attributed to a low-frequency $E_g$ Raman phonon[11,12]. Additionally, long-lived oscillations at GHz frequencies were attributed to the generation and propagation of strain waves[11].

Characterizing strain- or phonon-induced oscillations using reflectivity alone is challenging, as reflectivity primarily probes electronic changes and surface effects, offering limited insight into atomic-scale lattice dynamics. Here, we report time-resolved X-ray diffraction measurements on epitaxially strained $LaAlO_3$ thin films driven by THz-excited IR-active phonons. X-ray diffraction provides a direct measure of atomic displacements, and thin films offer an ideal platform: their geometry avoids restrictive grazing-incidence or grazing-exit geometries, the film-substrate interface influences nanoscale lattice dynamics and phonon scattering, and epitaxial strain engineering can modify vibrational properties. We observe an immediate lattice expansion and a breathing mode, explained by first-principles theory and a spring-mass model, confirming direct phonon-strain coupling. Optical birefringence shows the strain amplitude scales linearly with fluence, further validating the theory. Finally, analysis of the X-ray diffuse scattering—interpreted via a phase-field model—reveals enhanced crystallinity and structural ordering arising from domain boundary regions that expand laterally at the speed of sound.

*Photoinduced lattice expansion and crystalline ordering* – We study a 20 nm thick crystalline $LaAlO_3$ film epitaxially grown on a (001)-terminated $(NdAlO_3)_{0.39}(SrAl_{1/2}Ta_{1/2}O_3)_{0.61}$ (NSAT)[13,14] substrate by ozone-assisted molecular-beam epitaxy (see Supplementary Information). The pseudo-cubic [001] direction of $LaAlO_3$ is normal to the film surface, and epitaxy results in a 1.3% biaxial tensile strain (see Supplementary Information). $LaAlO_3$ has infrared (IR) active modes around 20 THz in the room temperature bulk rhombohedral phase. Our calculations predict that substrate clamping modifies the space group of $LaAlO_3$ (to orthorhombic, similar to Ref. [15], see Supplementary Information) and slightly changes the frequencies of the IR-active modes. Figure 1a shows the oscillator strength of IR phonons around 20 THz, as calculated from our first-principles theory for strained $LaAlO_3$.

Figure 1b shows the experimental setup for the time-resolved X-ray diffraction experiment conducted under ambient conditions. A mid-infrared laser pump and X-ray free-electron laser probe scheme measure the structural response of the strained $LaAlO_3$ film to resonant excitation of the IR-active modes. The mid-infrared pulses have a full width at half maximum (FWHM) duration of approximately 300 fs, a carrier frequency of 20 THz, a bandwidth of 4.7 THz [Fig. 1a], a fluence of 8.5 $mJ/cm^2$, and a spot size of 400 μm, with no carrier-envelope phase matching. The pump is p-polarized, so the electric field is parallel to the scattering plane. The X-ray pulses have a photon energy of 9.5 keV, a pulse duration of 10 fs, and a spot size of 200 μm. An area detector, with a pixel size of 75 μm, placed 600 mm downstream from the sample collects diffraction images.

We use the pseudo-cubic 008 Bragg peak of the $LaAlO_3$ film as a reporter of lattice dynamics. Data on the 00L and 02L Bragg peaks with L=10 show similar behavior. We make two central observations by analyzing the rocking scans measured 20 ps after photoexcitation. First, we observe a peak shift to lower momentum transfer, $q_∥$, indicating a lattice expansion in the direction parallel to the film-normal, [001] [Fig. 1c]. Second, we observe a decrease in diffuse scattering intensity and an increase in peak intensity, indicating light-induced ordering of the crystal planes [Fig. 1d].

Figure 2 shows the dynamics leading to the lattice expansion and ordering at 20 ps. We analyze three time-delay scans at different incident angles θ: at the angle of maximum intensity and angles on either side of the peak corresponding to half the maximum intensity. For each time delay, $\tau$, we fit the three intensities with $I(q,\tau) = I_P(\tau) \cdot \text{sinc}([q - q_P(\tau)]/W)^2$. Here, $I_P(\tau)$ is the peak intensity, $q$ is momentum transfer, $q_P(\tau)$ is the peak position, and $W$ is the peak width along the [008] direction fixed to the value found from the rocking scans. From the fit, we determine the time-dependent peak shift parallel to the film normal (indicating the lattice expansion) and the relative increase in the peak intensity (indicating the crystalline ordering).

The lattice expansion, $\Delta c(\tau)/c_0 \approx \Delta q_P(\tau)/q_{P,0}$ (subscript *0* denotes the ground state) begins immediately after photoexcitation, indicative of impulsive excitation[16]. The expansion saturates following an exponential decay, $\Delta c(\tau)/c_0 = a - b \cdot \exp(-\tau/\tau_d)$, with $a$ =0.0338(3)%, $b$ =0.0364(8)%, and a decay time of $\tau_d$=4.1(0.2) ps [Fig. 2a]. A strongly damped oscillation with a period of 3.7(2) ps appears when subtracting the fit from the data [Fig. S1]. After 237 ps, the lattice expansion, 0.042(4)%, exceeds the saturation reproduced by the exponential fit to the early-time data. The additional increase likely arises from the thermalization of the mechanical energy in the phonons. The peak intensity, $\Delta I_P(\tau)/I_{P,0}$, shows a different time behavior. Initially, the intensity marginally decreases, likely due to a change in the Debye-Waller factor[17] as the THz pump couples to the IR-active modes. After 5 ps, the intensity increases steadily, indicating an enhancement in crystalline ordering, and reaches a value of $\Delta I_P(\tau)/I_{P,0}$ =0.05 at 20 ps. The ordering decreases between 20 ps and 237 ps, in direct contrast to the increase of the lattice expansion.

Figure 2b displays the lattice ordering against the lattice expansion after photoexcitation. As expected from Figure 2a, the curve is nonlinear in the sub-20 ps range, and the data at 237 ps diverge from the short-time trend. Quasistatic measurements show that heating induces approximately linear expansion of the c-lattice and a linear increase in peak intensity between room temperature and 600 K [Fig. S2]. We also observe a peak sharpening, indicating an improved crystalline order. At 373 K, the ratio $\Delta I_P/\Delta c$ is about ten times smaller than in the photoinduced state after 20 ps. The difference between the dynamic and quasistatic data indicates that THz light induces a non-equilibrium structural state characterized by relatively lower lattice expansion and higher lattice ordering. Given that the optical phonons have lifetimes of a few picoseconds, the observed state indicates a metastable ordering after the coherent excitation of the lattice.

*Theoretical description of the lattice expansion* – We develop a phenomenological theory of the lattice dynamics by expanding the lattice energy, $U_{lat}$, of the crystal in terms of phonon amplitudes and *c*-axis strain, similar to our previous work[18]. Our lattice energy includes all symmetry-allowed couplings up to third order (see End Matter), and the values for the various terms that appear in the expansion are calculated from first-principles density functional theory. We numerically solve the differential equations of motion of $U_{lat}$ (see Supplementary Information) with respect to strain and phonons, where both degrees of freedom were initialized to zero amplitude; quantitative agreement of the induced lattice expansion with the experiment was achieved when we used an electric field amplitude that was 90% of the experimental peak electric field in our equations of motion simulations [Fig. S3a].

Figure 3a shows the calculated time-dependent *c*-axis strain, $\epsilon$, based on a model that includes the IR phonons and their coupling with strain. Our theory predicts that three IR phonons, each with a different symmetry, have frequencies near the carrier frequency of the pump pulse [Fig. 1a]. The IR phonon polarized along the *c*-axis contributes the largest change to the *c*-axis strain. The orange line in Figure 3a represents the lattice expansion calculated as a sum of contributions from all three high-frequency IR-active phonons. Our model suggests that the three IR modes and their coupling to strain contribute predominantly to the lattice expansion induced by mid-infrared excitation. We estimate the average expansion of the *c*-axis for a single unit cell yielding 0.027% for IR modes only and 0.032% for the full phonon spectrum (see End Matter). These estimates are consistent with the time-dependent calculation shown in Figures 3a,d.

The unit cell [Fig. 3a] expands significantly faster than the formation of coherent acoustic phonons that carry the expansion of the film [Fig. 2a]. To describe the film expansion, we use a classical one-dimensional spring-mass model (see Supplementary Information). We represent each unit cell by its mass connected to neighboring masses via springs with a strength equivalent to $c_{33}$, the elastic constant for stretching the *c*-direction [Fig. 3b]. The time-dependent strain from the unit cell simulations serves as the instantaneous equilibrium distance between the masses. Numerically solving the differential equation describing the interactions between unit cells yields the time-dependent atomic displacement for each mass, $u(x, \tau)$, which generates the local lattice constant by differentiation $c(x, \tau) = du(x, \tau)/dx$. The average of this local lattice constant is measured in the X-ray experiment. We assume the film (25 unit cells) is uniformly excited (the penetration depth of the pulse is on the order of hundreds of nanometers) and include atoms in the substrate (200 unit cells) to account for the inability of the film to expand freely into the substrate. We include an expansion of the substrate into the model because Fourier-transform infrared spectroscopy (FTIR) reveals the presence of IR phonons in the NSAT substrate with frequencies similar to the pumped LaAlO$_3$ modes [Fig. S5].

Figure 3c shows the change in the local lattice constant $\Delta c(x, \tau)$ as a function of time and the position in the film. Initially, the top layer of the film expands freely, while the bottom layer shrinks due to a larger expansion of the substrate. Subsequently, the local lattice constant oscillates due to acoustic phonons launched from the film-air and film-substrate interfaces. Figure 3d shows the time-dependent Bragg peak position determined as the position of the maximum of $I(q, \tau) = |\sum_j e^{iqx_j(\tau)}|^2$, where $x_j(\tau) = j \cdot c + u$, represent the time-dependent positions of atomic layers in the film. Combining unit cell dynamics with the spring-mass model reproduces the experimental data [Fig. 3d]: the calculated lattice expansion begins immediately after photoexcitation and saturates after a few ps. The saturation value depends linearly on the fluence [Fig. S4].

*Optical reflectivity and fluence dependence* – To provide further insight into the THz-induced dynamics, we conduct time-resolved THz pump-optical probe measurements, which are sensitive to Raman-like changes in the optical anisotropy [Fig. 4]. We measure the same films as the X-ray measurements, with similar pump characteristics; however, the pump pulse is incident normally on the film with only an in-plane polarization [Fig. 1a]. The time-resolved signal [Fig. 4b] shows fast coherent oscillations at early times and a slow frequency background signal at long times. A Fourier transform of the signal indicates the presence of several coherent modes excited by the pump pulse, most notably the $B_{3g}$ Raman mode (~1.2 THz) and a longitudinal acoustic mode arising from THz-induced strain (~0.3 THz). This ~0.3 THz frequency aligns with the 3.7 ps breathing mode observed in the X-ray experiment and predicted by our model. We further find that the strain-related signal arising at long times is nearly polarization-independent [Fig. 4d] and is linear in the fluence of the incident pulse (or quadratic in the peak electric field), corroborating our theoretical calculations, in which the out-of-plane strain induced by two different IR-phonons with in-plane polarization is approximately identical.

*Dynamics in the diffuse scattering* – Figure 5a shows that the diffuse scattering intensity decreases at different times for different $\Delta q_\perp$. To understand the origin of the diffuse scattering dynamics, we first consider the role of defects of structural order in a general framework. We model the time-resolved data by assuming identical, non-interacting defects of structural order[17, 19, 20] whose centers remain fixed after photoexcitation. The model does not depend on the type or distribution of structural defects (see End Matter). Yet, it shows (see Supplementary Information) that a disturbance originates at each defect and grows laterally [Fig. 5b] with a speed of 4.1(3) nm/ps [Fig. 5c], consistent with the transverse acoustic wave speed in LaAlO$_3$[11, 21].

Static diffuse scattering profiles, where the logarithm of scattering intensity is linear in $\Delta q_\perp$, have been reported in systems featuring partially ordered misfit dislocations[22-24]. Such dislocations arise from strain relaxation, but reciprocal space maps reveal that our films are commensurately strained [Fig. S8], ruling out misfit dislocations. Instead, in bulk LaAlO$_3$ at room temperature, domains with different octahedral rotations coexist[9, 25, 26], separated by boundaries where the rotations decrease. This reduction in rotation leads to a few % increase in the unit cell structure factor [Fig. S9]. Solving the Allen-Cahn equation[27] with octahedral rotations as the order parameter[9], we closely reproduce the observed diffuse scattering [Fig. 5d]. Heating weakens this scattering [Fig. S4], consistent with a transition towards the cubic structure in bulk[9]. These results support a scenario in which twinned domains, defined by differing octahedral rotations, exist in the ground state, and upon photoexcitation, the more symmetric boundary region expands into neighboring domains at the speed of sound, driven by a propagating sharp boundary.

*Discussion* – Our experiments demonstrate that THz-driven IR-active phonons in LaAlO$_3$ thin films induce an immediate lattice expansion and launch a strain-driven breathing mode. By combining time-resolved X-ray diffraction and optical birefringence measurements—supported by first-principles theory and a spring-mass model—we confirm the direct coupling between IR-active phonons and strain and validate that the strain amplitude scales linearly with the applied field. Beyond these core findings, analysis of time-resolved diffuse X-ray scattering indicates that the THz excitation interacts with preexisting domain walls, where octahedral rotations are reduced. This more symmetric structure expands ballistically into neighboring domains at the speed of sound. Unlike X-ray measurements of phonon dispersions in homogeneous materials[28-30], our approach reveals how defects serve as dynamic scattering centers that shape phonon propagation and structural evolution. In synergy with theory, our multimodal approach opens pathways to explore a broader range of lattice distortions and ultrafast lattice control strategies in complex oxides and heterostructures.

*Acknowledgments* – This work was primarily supported by the Department of Energy – Office of Basic Energy Sciences under award DE-SC0019414 (J.Z.K. and N.A.B. for first-principles theory; Y.E.S. and D.G.S. for thin film synthesis; R.S.R and J.W.H. for FTIR measurements; S. B., C.Y., J. H., and A.S.D. for THz reflectivity measurements; J.G., O.Y.G., and A.S for time-resolved X-ray measurements; Z.S and Y.A.B for collecting reciprocal space maps, R.C.F and A.S. for phase-field modeling and interpretation of diffuse X-ray scattering). G.K. was supported by the Cornell Center for Materials Research with funding from the NSF MRSEC program (Grant No. DMR-1719875). Computational resources were provided by the Cornell Center for Advanced Computing.

**Figures:**

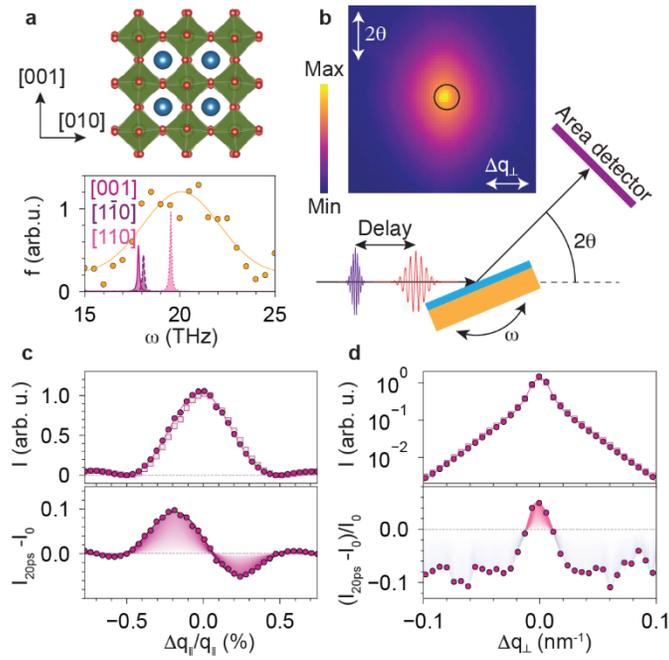

**Figure 1: THz-light induced strain and crystalline ordering.** (a) The crystal structure of epitaxially strained LaAlO$_3$ viewed along the [$\bar{1}$00] direction and the oscillation strength, $f$, of the infrared-active phonon modes with the polarization along [1$\bar{1}$0] (solid line), [001] (dashed line), and [110] (dotted line). The spectrum of the THz pump pulse used in the X-ray diffraction experiment is shown with orange circles and a Gaussian fit. (b) Pump-probe experimental setup: a p-polarized mid-infrared pulse excites a crystalline LaAlO$_3$ film. A colinear, delayed X-ray pulse probes the photoexcited state through diffraction at around $\theta = 43.85°$, collected on an area detector. A typical diffraction image shown in logarithmic scale spans 0.43 degrees in 2$\theta$. (c, d) The diffraction intensity of the 008 Bragg peak parallel (c) and perpendicular (d) to the Bragg rod and its difference from the ground state recorded 20 ps after photoexcitation.

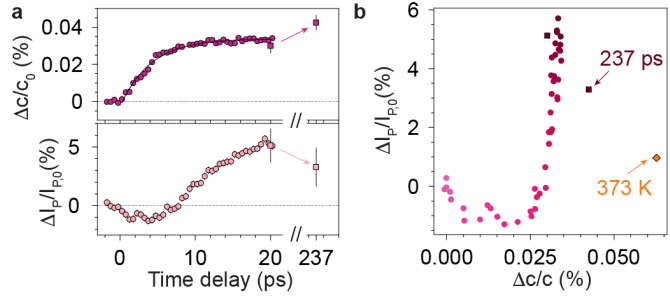

**Figure 2: Short-time dynamics after THz-excitation.** (a) Light-induced changes of the *c*-lattice parameter and exponential fit (top) and of the Bragg peak intensity (bottom). Note the lattice expansion increases and the peak intensity deceases from 20 ps to 237 ps (arrows). (b) The change in the peak intensity plotted against the change in the lattice constant (color represents time delay from 0 ps (light) to 20 ps (dark)). The square shows the change due to heating from room temperature to 373 K [Fig. S2]. Data extracted from time-delay scans at three incident angles (circles) and rocking scans at fixed time delays (squares) are shown: all measurements are consistent at 20 ps. Uncertainties in (a) (vertical lines if larger than the symbol) show the variation of the X-ray signal in the ground state (circles) or deviation between two separate rocking scans (squares).

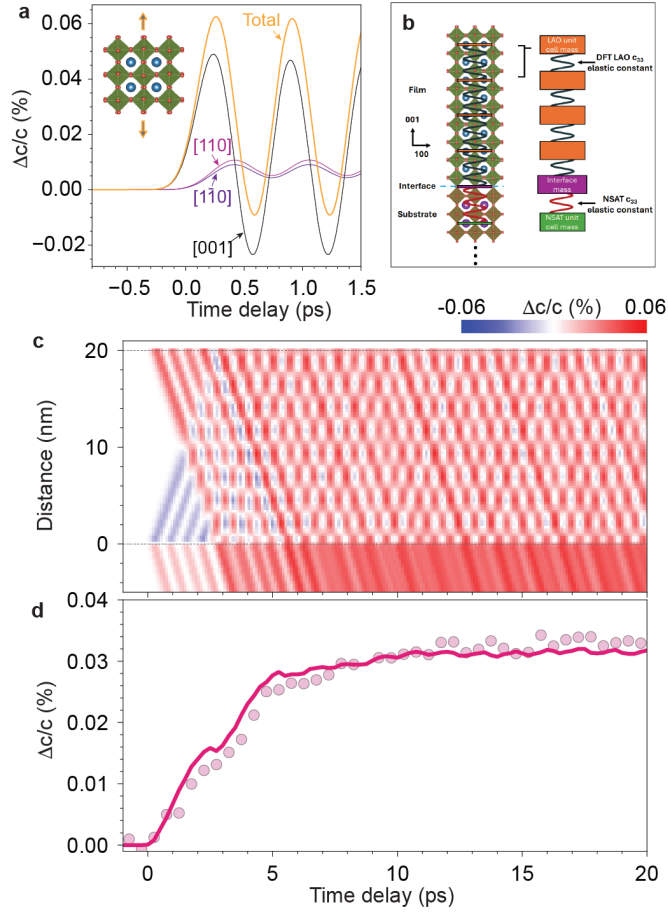

**Figure 3: Theoretically calculated response to THz-excitation.** (a) Simulations of the *c*-axis change in a single unit cell induced by the THz-pulse for the contributions of the field along [1$\bar{1}$0] (violet), [110] (magenta), and [001] (black), where only the coupling between IR modes to strain were considered. Each line shows the calculation where the electric field (oriented consistent with the experimental geometry [Fig. 1b]) couples to the single IR-active mode along that direction. The summation of the individual excitations (orange). (b) Schematic of the spring-mass model. (c) The local change in the lattice constant in the film. The substrate-film interface is positioned at 0 nm and the film is 20 nm thick. (d) The Bragg peak position calculated from the time and position dependent *c*-lattice constant shown in (c) (red line) and the experimental results reproduced from Figure 2(a) (shaded circles).

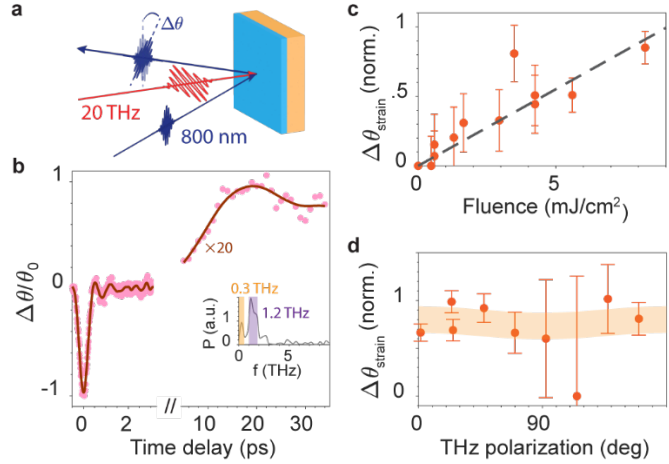

**Figure 4: Ultrafast optical measurements of THz-induced dynamics.** (a) Schematic of the experimental THz pump-optical probe setup. The pump pulse is incident normally and the optical probe comes at an angle of ~10°. The polarization rotation of the reflected probe pulse is detected as a function of time delay. (b) Time-resolved polarization rotation, normalized to the signal at $\tau = 0$, shown for both early and late times. Inset shows the Fourier transform of the signal indicating two strong peaks at 1.2 THz ($B_{3g}$ mode) and ~0.3 THz (acoustic mode). (c) Fluence dependence and (d) THz polarization dependence of strain-related signal at long time delays ($\tau=40$ ps).

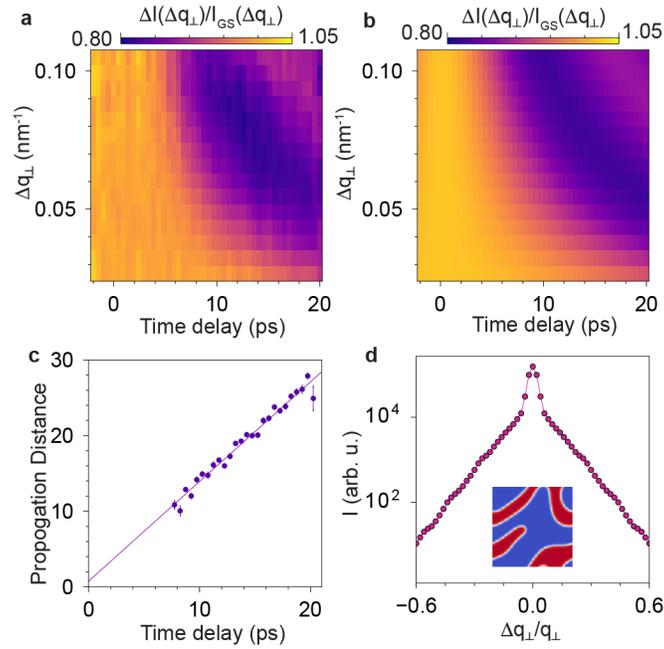

**Figure 5: THz-induced dynamics of the in-plane heterogeneity.** (a) Normalized change in the diffuse scattering intensity as a function of momentum transfer and time delay. (b) Fit of the data using the model (see Supplementary Information). (c) Propagation distance plotted against time delay and a linear fit; data at delays shorter than 7.5 ps are omitted due to uncertainties exceeding the displayed range. (d) Diffuse scattering simulated using the Allen-Cahn model. Inset: A typical spatial distribution of domains is shown in false color, with the domain boundaries highlighted in white.

**End Matter**

**Phenomenological Model of Crystal Lattice Energy**

We model the lattice dynamics by writing the lattice energy, $U_{lat}$, of the unit cell including all symmetry-allowed couplings up to third order:

$$U_{lat} = \tfrac{1}{2}\sum_i K_{i,R} Q_{i,R}^2 + \tfrac{1}{2}\sum_i K_{i,IR} Q_{i,IR}^2 + \tfrac{1}{2} K_\epsilon \epsilon^2 + \sum_{ijk} A_{ijk} Q_{i,R} Q_{j,IR} Q_{k,IR} +$$
$$\sum_{ijk} B_{ijk} Q_{i,\Gamma_1^+} Q_{j,\Gamma_1^+} Q_{k,\Gamma_1^+} + \sum_i C_i \epsilon Q_i^2 + D\epsilon^3 + \sum_{ij} E_{ij} Q_{i,\Gamma_1^+} Q_{j,\Gamma_1^+} \epsilon + \sum_i F_i Q_{i,\Gamma_1^+} \epsilon^2 +$$
$$\sum_i G_i Q_{i,\Gamma_1^+} \epsilon - \sum_i \tilde{Z}_i^* \vec{E} Q_{i,IR} .$$

Here, the $Q$'s [Å] represent phonon amplitudes for Raman-active ($Q_R$) and IR-active ($Q_{IR}$) phonons, $\epsilon$ is the c-axis strain (which has $\Gamma_1^+$ symmetry), and the $K$'s [eV/Å²] are the respective second-order force constants. The summations over $i, j,$ and/or $k$ are carried out over each phonon of its relevant character, i.e., $Q_{i,IR}$ covers every IR-active phonon, while $Q_{i,\Gamma_1^+}$ describes phonons of $\Gamma_1^+$ character. Terms starting with letters $A$ to $F$ [eV/Å³] represent third-order lattice anharmonicities, where the leading coefficient is the coupling strength. $G_i$ [eV/Å²] is a bilinear coupling coefficient between $\Gamma_1^+$ modes and strain. $\tilde{Z}_i^*$ [e⁻] is the mode-effective-charge, describing the coupling between an incident electric field, $\vec{E} = \eta E_0 f \tau$, and the infrared active phonons. Here, $E_0$ [MV/cm] is the peak electric field, $\tau$ [ps] is the duration of the field assuming a Gaussian envelope, $\eta$ is a dimensionless parameter determined by the shape of the pulse (for a Gaussian $\eta \approx 1.67$), and $f$ is the linear carrier-frequency of the pulse, which we set to be the average frequency of the IR phonons polarized parallel to the substrate (17.96 THz). The values for the different $K$, $A$-$G$, and $\tilde{Z}^*$ were calculated from first-principles density functional theory. Using techniques similar to Ref.[18], the analytical form of the time-averaged induced lattice expansion due to the IR-active phonons can be written in terms of parameters in $U_{lat}$: $\langle \epsilon_{33} \rangle = \sum_i \frac{2 C_i \tilde{Z}_i^{*2} f_{IR}^2}{c_{33} K_{i,IR}^2} (\eta E_0 \tau)^2$.

**Model for the Time-Resolved Diffuse X-ray Scattering**

We assume that a defect modifies the local unit cell structure factor, $F(\mathbf{Q})$ (a similar result emerges when one considers small displacements around defects). Assuming identical and spatially distributed defects[19], we use $\phi(\mathbf{r})$ to represent the spatial dependence of the defect-induced change in the structure factor, where $r = R_s - R_t$ is the distance from a defect located at $R_t$ to a particular unit cell at $R_s$. The total scattering amplitude is then:

$$E(Q) = \sum_s \left[ F(Q) + \sum_t \phi(R_s - R_t) \right] e^{iQ \cdot R_s}.$$

The first term in the brackets, $\sum_s F(Q) e^{iQ \cdot R_s}$, gives rise to Bragg peaks at reciprocal lattice points, $G$. The second term, $\sum_s \sum_t \phi(R_s - R_t) e^{iQ \cdot R_s}$, introduces diffuse scattering due to the defects. When the interference between Bragg and diffuse scattering is negligible, one can write $I(Q) = I_B(Q) + I_D(Q)$. The diffuse scattering intensity measured around each reciprocal lattice point $G$ is then

$$I_D(\Delta q) = |\Phi(\Delta q)|^2 \left| \sum_t e^{i(G + \Delta q) \cdot R_t} \right|^2,$$

where $\Phi(\Delta q)$ is the continuous Fourier transform of $\phi(r)$, $\Phi(\Delta q) = \int_{-\infty}^{\infty} \phi(r) e^{-i\Delta q \cdot r} \, dr$. Here, $\Delta q = Q - G$, measures the deviation from the Bragg condition. If defects reside at lattice sites, $e^{iG \cdot R_t} = 1$ and $I_D(\Delta q) = |\Phi(\Delta q)|^2 \left| \sum_t e^{i\Delta q \cdot R_t} \right|^2$. Assuming photoexcitation modifies the unit cell structure factor around the defect, $\phi_0(r)$ to $\phi(r, \tau)$, yet the positions of the defects, $R_t$, remain unchanged, we can calculate the dynamics of each single, identical defect via

$$\frac{I_D(\Delta q, \tau)}{I_{D,0}(\Delta q)} = \frac{|\Phi(\Delta q, \tau)|^2}{|\Phi_0(\Delta q)|^2}.$$

The sum over defect positions cancels out, and the time-resolved diffuse scattering data can be analyzed with no knowledge of the defect distribution.

# Picosecond expansion in LaAlO$_3$ resonantly driven by infrared-active phonons

Jakob Gollwitzer,[1,*] Jeffrey Z. Kaaret,[2,*] Y. Eren Suyolcu,[1] Guru Khalsa,[1,3] Rylan C. Fernandes,[1] Oleg Gorobtsov,[1] Sören Buchenau,[2] ChanJu You,[4] Jayanti Higgins,[2] Ryan S. Russell,[5] Ziming Shao[1], Yorick A. Birkhölzer[1], Takahiro Sato,[6] Matthieu Chollet,[6] Giacomo Coslovich,[6] Mario Brützam,[7] Christo Guguschev,[7] John W. Harter,[5] Ankit S. Disa,[2] Darrell G. Schlom,[1,7,8] Nicole A. Benedek,[1, †] and Andrej Singer[1, ‡]

[1]*Department of Materials Science and Engineering, Cornell University, Ithaca, New York 14853, USA2*
[2]*School of Applied and Engineering Physics, Cornell University, Ithaca, New York 14853, USA*
[3]*Department of Physics, University of North Texas, Denton, Texas 76203, USA*
[4]*Department of Physics, Cornell University, Ithaca, NY 14853, USA*
[5]*Materials Department, University of California, Santa Barbara, California 93106, USA*
[6]*The Linac Coherent Light Source, SLAC National Accelerator Laboratory, Menlo Park, CA 94025 USA*
[7]*Leibniz-Institut für Kristallzüchtung, Max-Born-Straße 2, 12489 Berlin, Germany*
[8]*Kavli Institute at Cornell for Nanoscale Science, Cornell University, Ithaca, NY 14853, USA*
*These authors contributed equally to this work.
[†]*nbenedek@cornell.edu*, [‡]*asinger@cornell.edu*

**Thin film growth**

LaAlO$_3$ films were all grown by ozone-assisted molecular-beam epitaxy[1, 2] on (001)-terminated (NdAlO$_3$)$_{0.39}$(SrAl$_{1/2}$Ta$_{1/2}$O$_3$)$_{0.61}$ (NSAT) substrates at a growth temperature of 950 °C. The growth temperature is measured by using a thermocouple that is positioned close to but not in direct contact with the substrate; the actual temperature of the substrate is roughly 850 °C. During the growth, distilled ozone – a mixture of approximately 80% ozone and 20% oxygen – was introduced at a background partial pressure of $1 \times 10^{-6}$ Torr. The fluxes of lanthanum and aluminium, evaporated from MBE effusion cells, were first adjusted to achieve fluxes of ~$1.3 \times 10^{13}$ atoms cm$^{-2}$s$^{-1}$, as initially determined by a quartz crystal microbalance. The fluxes were then more precisely calibrated *(i)* by measuring the thickness of a lanthanum oxide calibration film grown on a (111) (ZrO$_2$)$_{0.905}$(Y$_2$O$_3$)$_{0.095}$ substrate by X-ray reflectivity (XRR), and *(ii)* by utilizing the monitored RHEED oscillations during the growth Al$_2$O$_3$ thin films on ($1\bar{1}02$) Al$_2$O$_3$ substrates. Further details of such flux calibration can be found in Ref. [3]

**Experimental Details**

We performed two types of X-ray measurements. A detector image at a fixed incident angle θ measures a slice in the reciprocal space along the Ewald sphere. We measured these slices at different angles via rocking the sample for collecting the 3D reciprocal space around each Bragg peak at time delays of 2.5 ps, 20 ps, and 237 ps. These data allow us to track the shift of the Bragg peak, inversely proportional to lattice constant changes, and the peak width, inversely proportional to the structural correlation length. We complemented the rocking data with measurements taken at a fixed incident angle: at the Bragg condition, as well as above and below it, all with a temporal resolution within 1 ps. Combined with the rocking scan, we related the intensity on a fixed slice of the Ewald sphere to the peak shift. We also extracted the peak width perpendicular to the q vector from the measurement at a fixed θ.

**Fourier-Transform Infrared Absorption**

Fourier-transform infrared spectroscopy (FTIR) measurements were performed on the NSAT/LaAlO$_3$ sample. The measurements were performed at room temperature with unpolarized IR light using attenuated total reflectance with diamond. Figure S5a shows the absorbance for a bulk LaAlO$_3$ substrate. Figure S5b shows the absorbance of a NSAT (100) substrate, and the NSAT/LaAlO$_3$ sample system. The LaAlO$_3$ substrate absorbs strongly at 19.1 THz. The NSAT substrate also shows absorption peaks near that frequency. The absorbance of the NSAT/LaAlO$_3$ sample system is very similar to that of the NSAT substrate, which is to be expected for a thin film as the evanescent wave passes through the film and extends into the NSAT substrate. Figure S5c show the NSAT substrate absorbance subtracted from the NSAT/LaAlO$_3$ absorbance. This difference absorbance roughly mimics the bare LaAlO$_3$ absorbance. We may take this difference absorbance to be the qualitative absorbance of the film itself. A broad absorption band is observed around this main peak for the sample system. Unfortunately, as these measurements were performed at room temperature, we cannot distinguish individual phonon modes within this broader spectrum.

**Symmetry of the Film**

Above 813 K bulk LaAlO$_3$ adopts the 5-atom cubic perovskite structure ($Pm\bar{3}m$, space group #221). With cooling it undergoes a structural phase transition to a rhombohedral structure due to the condensation of out-of-phase octahedral tilts about the [111] axis ($R\bar{3}c$, space group #167)[4]. In Glazer notation the octahedral tilt pattern is denoted as $a^-a^-a^-$, where the $a$'s represent equal amplitude rotations about the $a$, $b$, and $c$ axes and the $^-$ designate that the rotations are out-of-phase along each axis[5, 6].

Epitaxial strain can modify the octahedral rotation pattern in LaAlO$_3$. Tensile strain induces equal-amplitude out-of-phase tilts about the in-plane axes ($a^-a^-c^0$), altering the crystallographic symmetry to orthorhombic (*Imma*, space group #74). Compressive strain induces an out-of-phase tilt about the surface normal ($a^0a^0c^-$), making the system tetragonal (*I4/mcm*, space group #140)[7]. Since NSAT imposes tensile strain, we assume the LaAlO$_3$ thin-film is in the *Imma* phase[8], resulting in 27 optical phonons, which transform like the following irreducible representations:

$$\Gamma_{\text{optic}} = 2\Gamma_1^- \oplus 5\Gamma_2^- \oplus 4\Gamma_3^- \oplus 4\Gamma_4^-$$
$$\oplus 3\Gamma_1^+ \oplus 2\Gamma_2^+ \oplus 3\Gamma_3^+ \oplus 4\Gamma_4^+$$

**Computational Details**

Phonon-phonon and phonon-strain coupling were calculated using density functional theory (DFT), as implemented in VASP 6.2.0 [9, 10], using the projector augmented-wave (PAW) method [11] and the PBEsol exchange-correlation functional[12]. The following states were included in the valence of the PAW potentials: $5s^25p^65d^16s^2$ for La, $3s^23p^1$ for Al, and $2s^22p^4$ for O. A force convergence tolerance of $10^{-4}$ eV/Å was used for all calculations with a 4×4×4 Monkhorst-Pack *k*-point grid and plane-wave energy cutoff of 600 eV in a 40-atom supercell. These values were chosen to converge phonon frequencies, calculated with density functional perturbation theory [13] (DFPT), to within 5 cm$^{-1}$ when compared to incrementing the *k*-point grid to 12×12×12, and the energy cutoff to 1300 eV. Previous work has shown that the nominally empty *f* states of the lanthanum atom are too low in energy which leads to spurious mixing with other states [14]. To correct for this, we used a value for bulk LaAlO$_3$ of $U - J =$ 10.32 eV [8] for the on-site Coulomb interaction for the lanthanum *f* orbitals [15].

Bulk LaAlO$_3$ is rhombohedral at room temperature with a psuedocubic lattice constant of 3.79 Å [4], while the NSAT substrate has a lattice constant of 3.84 Å [16, 17]. By defining the in-plane strain as $\epsilon = \frac{a_{sub} - a_{film}}{a_{film}}$, where $a_{\text{sub}}$ is the pseudocubic spacing of (unstrained) NSAT at room temperature and $a_{\text{film}}$ is the psuedocubic spacing of (unstrained) LaAlO$_3$ at room temperature, an effective in-plane strain of 1.3% is applied on the film. This strain was applied to our calculated $R\bar{3}c$ LaAlO$_3$ structure with psuedocubic lattice constants of 3.79 Å, which lowers the space group symmetry to *Imma*. Our converged lattice constants in the *Imma* phase (using the convergence criteria in the previous paragraph) are 7.696 Å for the *a* and *b* axes and 7.541 Å for the *c*-axis, with an out-of-phase tilt angle of 3.78° about each in-plane axis.

Second- and third-order coupling coefficients were calculated via a series of symmetry-constrained frozen-phonon calculations. Mode-effective charges were calculated as defined in Ref. [18], which were found to be consistent with values obtained from modern theory of polarization calculations [19-21]. All meshes went up to ±60 pm displacements of phonons in 3 pm increments and the *c*-axis strain up to ±4% in 0.2% increments. Symmetry assignment of the irreducible representations of phonon modes were generated with the ISOTROPY Software Suite [22,23]. Calculated phonon frequencies, reduced masses, and mode-effective-charges of these phonons from our first-principles simulations in the strained *Imma* structure can be found in Table S1.

**Table S1**: Phonon frequencies, reduced masses and mode effective charges (MEC) of all zone-center phonons in 1.19% tensile strained LaAlO$_3$ from our first-principles calculations.

| Mode Symmetry | Frequency [THz] | Mass [AMU] | MEC [e$^-$] |
|---|---|---|---|
| $\Gamma_1^-$ (A$_u$) | 13.96 | 26.88 | |
|  | 9.20 | 16.04 | |
| $\Gamma_2^-$ (B$_{1u}$) | 18.08 | 16.37 | 3.63 |
|  | 14.92 | 26.84 | 1.17 |
|  | 11.91 | 22.31 | 27.27 |
|  | 9.59 | 16.07 | 0.02 |
|  | 5.54 | 25.02 | 6.92 |
| $\Gamma_3^-$ (B$_{3u}$) | 19.52 | 16.88 | 5.63 |
|  | 13.92 | 26.98 | 0.30 |
|  | 12.49 | 21.21 | 0.95 |
|  | 5.33 | 25.37 | 30.71 |
| $\Gamma_4^-$ (B$_{2u}$) | 17.82 | 16.38 | 3.95 |
|  | 11.81 | 22.57 | 26.78 |
|  | 8.49 | 16.07 | 0.58 |
|  | 4.83 | 24.59 | 6.87 |
| $\Gamma_1^+$ (A$_g$) | 13.86 | 16.02 | |
|  | 4.70 | 27.68 | |
|  | 4.12 | 29.72 | |
| $\Gamma_2^+$ (B$_{1g}$) | 12.77 | 16.00 | |
|  | 1.51 | 16.00 | |
| $\Gamma_3^+$ (B$_{3g}$) | 13.46 | 16.06 | |
|  | 4.13 | 111.67 | |
|  | 1.02 | 16.41 | |
| $\Gamma_4^+$ (B$_{2g}$) | 21.93 | 16.00 | |
|  | 13.65 | 16.00 | |
|  | 13.36 | 16.04 | |
|  | 4.45 | 136.26 | |

The elastic constants calculated from first principles can be found in Table S2. We used two different techniques. We used the built-in finite difference approach (IBRION=6 and ISIF=3) implemented in VASP. The 'Mesh' calculations shown in Table S2 involved altering the lattice constants (up to ±0.5%, in 0.05% steps) and fitting an eight degree polynomial to the resulting energy landscape. The elastic constants were calculated for both fixed ions and relaxed ions with the altered lattice constants to see if it would meaningfully change the calculated longitudinal speed of sound parallel to the surface normal. According to our calculations, allowing the ions to relax increases the speed of sound by 1.3% relative to fixed ions.

**Table S2:** Results from our first-principles calculations for elastic constants (all in Voigt notation) of 1.3% tensile strained LaAlO$_3$ in kBar.

| Method | $c_{11}$ | $c_{22}$ | $c_{33}$ | $c_{44}$ | $c_{55}$ | $c_{66}$ | $c_{12}$ | $c_{13}$ | $c_{23}$ |
|---|---|---|---|---|---|---|---|---|---|
| VASP relaxed | 2653 | 2657 | 3386 | 1307 | 1268 | 1491 | 1555 | 1251 | 1256 |
| Mesh relaxed | 2620 | 2542 | 3420 | 1285 | 1286 | 1495 | 1571 | 1223 | 1257 |
| VASP fixed | 3229 | 3229 | 3443 | 1600 | 1564 | 1573 | 1277 | 1296 | 1296 |
| Mesh fixed | 3242 | 3245 | 3475 | 1585 | 1585 | 1576 | 1286 | 1304 | 1307 |

**Simulated Dynamics**

$$F_{Q_R} = M_R \ddot{Q}_R = \frac{-\delta U_{\text{lat}}}{\delta Q_R}$$
$$= -K_R Q_R - \sum_{jk} A_{Rjk} Q_{j,IR} Q_{k,IR} - \sum_{jk} B_{Rjk} Q_{j,\Gamma_1^+} Q_{k,\Gamma_1^+} - 2C_R \epsilon Q_R - \sum_j E_{Rj} Q_{j,\Gamma_1^+} \epsilon$$
$$- F_R \epsilon^2 - G_{R(\Gamma_1^+)} \epsilon$$

$$F_{Q_{IR}} = M_{IR} \ddot{Q}_{IR} = \frac{-\delta U_{\text{lat}}}{\delta Q_{IR}} = -K_{IR} Q_{IR} - \sum_{ij} A_{ijIR} Q_{i,R} Q_{j,IR} - 2C_{IR} \epsilon Q_{IR} + \tilde{Z}_{IR}^* E$$

$$F_\epsilon = M_\epsilon \ddot{\epsilon} = \frac{-\delta U_{\text{lat}}}{\delta \epsilon}$$
$$= -K_\epsilon \epsilon - \sum_i C_i Q_i^2 - 3D\epsilon^2 - \sum_{ij} E_{ij} Q_{i,\Gamma_1^+} Q_{j,\Gamma_1^+} - 2\sum_i F_i Q_{i,\Gamma_1^+} \epsilon - \sum_i G_i Q_{i,\Gamma_1^+}$$

We used a Runge-Kutta 5(4) method for numerical integration of the differential equations of motion in the equations above (which are found by taking partial derivatives of $U_{\text{lat}}$ with respect to different coordinates), as implemented in NumPy [24]. Two types of simulations were performed. First, we calculated the phonon and strain response to optical excitation assuming the crystal could respond uniformly. In these simulations a Gaussian electric field with a full width at half maximum of $\tau$ = 424 fs, a peak electric field of 4.62 MV/cm, and a resonant frequency of 17.96 THz (the average of the high frequency $\Gamma_2^-$ and $\Gamma_4^-$ modes). These values are consistent with those used in the experiment. Unfortunately, they caused our simulations to overestimate the lattice expansion.

Nevertheless, we did not include dielectric screening and Fresnel reflection conditions, which would decrease the effective electric field in the film. Using a value of 90% of the peak experimental electric field amplitude, we recovered results quantitatively similar to those seen in experiment (Fig. S3a).

The *c*-axis strain from the single unit cell dynamics was used as an input for the second set of simulations: the propagation of strain waves in the excited film using a 1-dimensional spring-mass model as described in the main text. In the spring-mass model, the mass on each spring was the mass of our computational unit cell (1711.04 u), and the force constant of each spring was found by taking $c_{33}$ (Table S2), multiplying by the cell volume, dividing by the *c*-axis lattice constant squared, and then converting to the appropriate units with a resulting value of 16.25 eV/Å². While a force constant of 26.25 eV/Å² and a mass of 1836.00 u were used for as parameters for substrate. The speed of sound in LaAlO$_3$, which was calculated from first principles by finding the elastic constant and solving $v_z = \sqrt{\frac{c_{33}}{\rho}}$, with a resulting speed of 7.296 nm/ps ($c_{33}$ = 3386 kBar, $\rho$ = 3.83 AMU/Å³). This speed is comparable to previous experimental work in bulk LaAlO$_3$ which found the speed of sound along the [001] direction to be 6.7 nm/ps [4, 25]). Using this speed, we estimate the propagation time from the top to the bottom of the film to be 2.74 ps.

When modeling without an expansion of the substrate the *c*-axis overshoots and reaches a maximum value after ≈2.7 ps, half of what was seen in experiment before saturating at 5 ps. This is because strain waves formed at the air/film interface and at the film/substrate interface and propagate toward each other, allowing the standing wave to form in the time it takes to go from the top to the bottom of the film. The FTIR results in Figure S5 indicate an IR absorption in the substrate at the same frequency, suggesting that the phonon driven lattice expansion mechanism is also occurring in NSAT. The model for the time trace of the *c*-lattice constant shows a kink near 2.7 ps and a broader shoulder after 5.5 ps. The broad shoulder was replicated when the elastic constant of the substrate was much larger than that of the film. Figure S3 shows how varying different parameters in the spring-mass model alters the dynamics of the film. We show the dependence of the induced strain on the peak electric field, IR-phonon damping, the ratio of the substrate expansion relative to the film, and the force constant of the substrate corresponding to $c_{33}$. The substrate elastic constant also affected where the kink occurred at 2.7 ps in Δ*c*: the stiffer the elastic constant, the lower in Δ*c* the kink would occur and the broader the shoulder at 5.5 ps would be (Fig. S3d). Additionally, the kink at 2.7 ps is affected by how much the substrate expands relative to the film: the larger the substrate expansion, the lower in Δ*c* the kink would occur (Fig. S3c). Fig. S4 shows how the *c*-axis expansion (which is an average of the data from Fig. S3a, from 10 ps to 20 ps) varies linearly as a function of fluence (which is proportional to $E^2$). The long-lived nature of the driven strain observed experimentally contrasts the short lifetimes of the IR-active phonons, comparable to the long-lived strain waves seen in Ref. [25]. Our simulations which included damping of the IR-phonons did not reflect what was seen in experiment, with noticeable deviation for even improbably small values of IR-phonon damping (Fig. S3b), perhaps indicating a phonon bottleneck that limits how much energy can leave the IR modes. This phonon bottleneck might be a result of the thin film limiting the Brillion zone as most wave vectors along the c-axis are not representable in a 20 nm film. The constrained Brillion zone would mean fewer pathways for energy to scatter, potentially enhancing the lifetime of phonons[26].

**Fitting the time-resolved diffuse scattering**

To analyze the measured time-resolved diffuse X-ray scattering, we consider a 1d-model. Before photoexcitation, $\phi_0(r)$ is localized at the defect site, modeled as $\phi_0(r) = \delta(r)$, where $\delta(r)$ is the Dirac delta function. After photoexcitation, a disturbance grows to a size, d,

$$\phi(r,\tau) = c_1 \delta(r) + c_2 \Pi\left(\frac{r}{2d}\right),$$

where $\Pi(r) = 1$ if $r < 1/2$ and 0 otherwise. The Fourier transform yields

$$\Phi(\Delta q, \tau) = c_1 + 2c_2 d \operatorname{sinc}\left(\frac{\Delta q d}{\pi}\right),$$

where $\operatorname{sinc}(q) = \sin(\pi q)/(\pi q)$. In the ground state, $\Phi_0(\Delta q) = 1$, and the normalized intensity change is

$$\frac{I_D(\Delta q, \tau)}{I_{D,0}(\Delta q)} = c_1^2 + 4c_1 c_2 d \operatorname{sinc}\left(\frac{\Delta q d}{\pi}\right) + 4c_2^2 d^2 \operatorname{sinc}\left(\frac{\Delta q d}{\pi}\right)^2.$$

Using the equation above, we fit[27] the normalized time-resolved diffuse X-ray scattering intensity at each time delay, $\tau$. We find approximately constant values for $c_1^2 = 0.86$ and $4c_1 c_2 d = 0.16$ (sinc² term is negligible). Plotting $d$ against $\tau$ yields linear relationship, $d = \tau \cdot v$, with $v = 4.1(3)$ nm/ps. For time delays shorter than 7.5 ps, the fit results displayed large uncertainties. At these early times the sinc function is indistinguishable from a constant within the measured range in $\Delta q$. Additionally, the limited $\Delta q$ range prevented determining a characteristic defect size in the ground state; here, static defects are consistent with Dirac delta functions.

**Allen-Cahn model for calculating domain distribution**

We applied phase field modeling to gain a clearer understanding of the defect distribution and its origin. We chose the octahedral rotations as our order parameter, which are present in the bulk at room temperature and disappear at higher temperatures as the structure transitions from rhombohedral to cubic. We assume that neighboring domains exhibit the same magnitude of these octahedral rotations, resulting in identical structure factors. Nevertheless, at the boundaries between domains where the rotation directions differ, the rotations are bound to decrease. Figure S9 illustrates that at the 008 Bragg peak, the unit cell structure factor increases by approximately 10% when octahedral rotations are reduced. This change is a possible a physical basis for the diffuse scattering described by $\phi(\mathbf{r})$.

To compute the domain distribution, we assume a continuous transformation as the material cools from the cubic to the rhombohedral phase (In the thin film the structural transition is likely from nearly tetragonal orthorhombic to more orthorhombic, which we assume has no substantial impact on this discussion.) Although four domain types are possible, we neglect this complexity—which could be addressed with more sophisticated phase-field models[28]—arguing that it will not significantly change the overall distribution. To determine the domain distribution and find the positions of defects, $\mathbf{R_t}$, we solve the Allen-Cahn equation

$$\frac{\partial \eta}{\partial t} = -M\left(\frac{\partial f_{\text{hom}}}{\partial \eta} - 2K\nabla^2 \eta\right),$$

where $\eta$ is the order parameter, $M$ is the mobility, $f_{\text{hom}}$ is the homogeneous free energy, and $K$ is the gradient energy coefficient[29]. We use a double well potential for the homogeneous free energy $f_{\text{hom}} = f_{\text{max}}(1 - \eta^2)^2$. We initialize $\eta$ to zero and add Gaussian noise. The parameters are set to M=0.05, $f_{\text{max}}$=1.0, and K=2.25. We implement the equation in two dimensions using a finite difference method with Δx=Δy=1.0 and Δt=0.01 and ran it for 50,000 timesteps on a 200x200 grid [see inset in Fig. 5d]. After obtaining the final domain structure, we compute the gradient magnitude to identify the domain boundaries. The diffuse scattering is then calculated from the intensity of the two-dimensional Fourier transform of the gradient field. If defects are located at precisely at lattice sites, $e^{i\mathbf{G}\cdot\mathbf{R}_t} = 1$ and $I_D(\Delta\mathbf{q}) = |\Phi(\Delta\mathbf{q})|^2 \left|\sum_t e^{i\Delta\mathbf{q}\cdot\mathbf{R}_t}\right|^2$. The equation is the discrete Fourier transform we calculate from the domain boundaries located at $\mathbf{R_t}$. Figure 5d of the main text shows the central line scan of the 2D discrete Fourier transform of the gradient matrix and then average the intensity values over 50 simulations.

**SUPPLEMENTARY FIGURES**

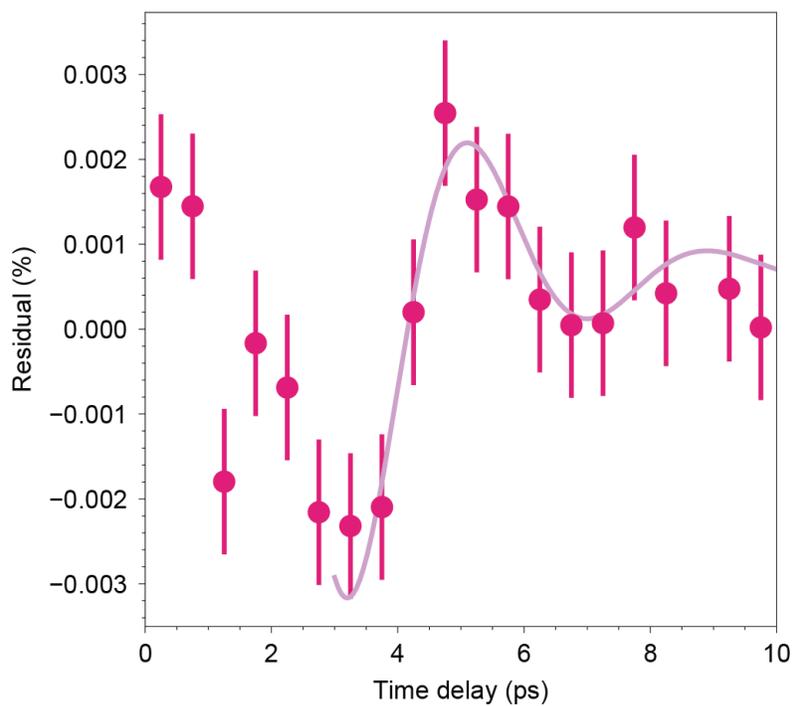

**Figure S1:** The residual obtained from the data shown in Figure 2a (top) after subtracting the exponential fit that is described in the main text. The residual is fit with a damped sinusoidal function in the range between 2.5 ps and 10 ps. Including data in the interval from 0 ps to 2.5 ps worsens the fit significantly. The period from the shown fit is 3.7(2) ps.

10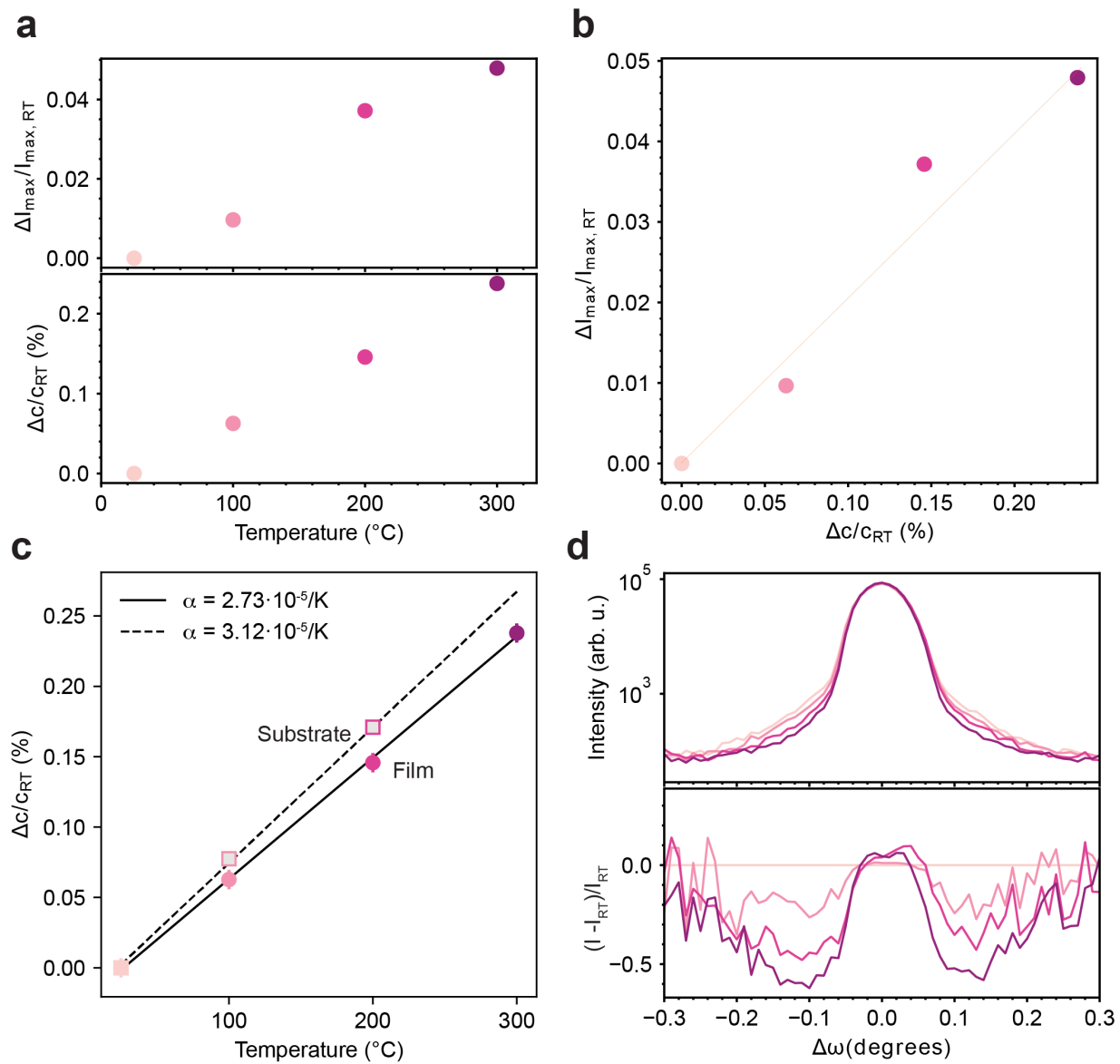

**Figure S2:** Quasistatic measurements of the LaAlO$_3$ thin films at various temperatures. (a) The increase in intensity (top) and lattice expansion (bottom) of the film as a function of temperature. (b) Replotting (a) where the intensity is a function of the lattice expansion. (c) The change in the lattice spacing of the substrate and film as a function of temperature, the thermal expansion coefficients are estimated from the linear fit. (d) Top: The Bragg peak at various temperatures, as temperature increases the Bragg peak narrows. Bottom: Subtracting the room temperature intensity from the various temperatures, showing the change in the diffuse scattering as a function of the rocking angle.

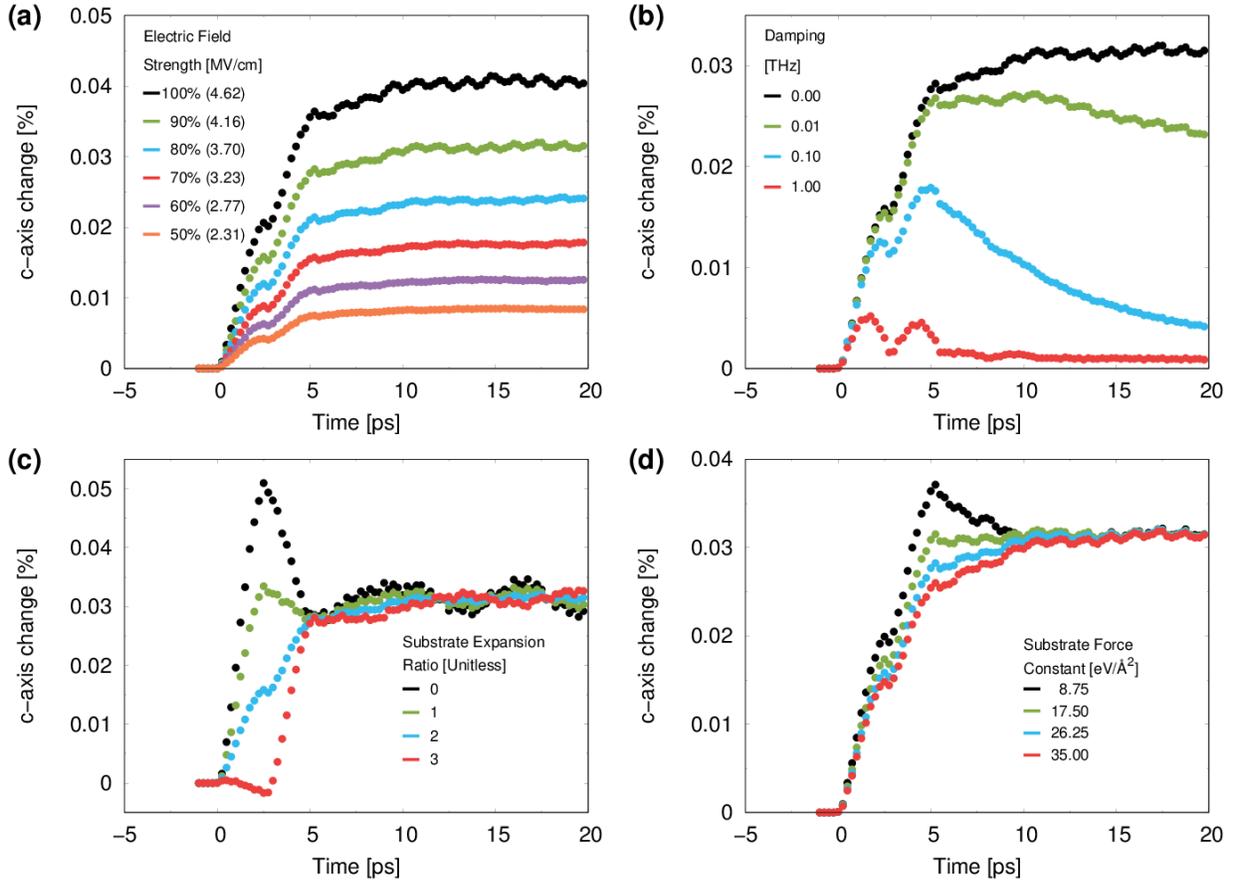

**Figure S3:** The dependence of the *c*-axis expansion in the 1-dimensional spring-mass model (the y-axis in each plot corresponded to a reciprocal lattice vector $q$ which maximized: $I(q,\tau) = |\sum_j e^{iqx_j(\tau)}|^2$) for various input parameters: (a) Dependence of the strain on electric field, where 100 % corresponds to the maximum experimental value of 4.62 MV/cm for the pump pulse. The numbers in the legend show electric field values in MV/cm scaled to the maximum experimental value. The induced *c*-axis length appears to scale quadratically with the electric field (Fig. S4 shows the c-axis expansion versus fluence). The oscillations about the new equilibrium seen for the larger fields are likely due to lattice anharmonicities contributing more to the strain response. (b) Dependence of the strain for various damping values of the infrared-active phonons. Even unreasonably small values of damping (very long lifetimes) damp the strain in ways not observed in the experiment. (c) Strain dependence of the film for different amounts of expansion in the substrate, by which we mean the relative expansion of the substrate compared to the film. For no substrate expansion and equal-film-substrate-expansion the strain in the film overshoots before returning to the meta-stable strain. If the substrate expands more than the film, the divot at 2.7 ps lowers and can even become negative. (d) Dependence of the strain for various values of the elastic force constant in the substrate (which we cannot model with DFT), force constants less than the film overshoot the meta-stable strain, whereas force constants greater than the film develop the broad shoulder between 5 and 10 picoseconds.

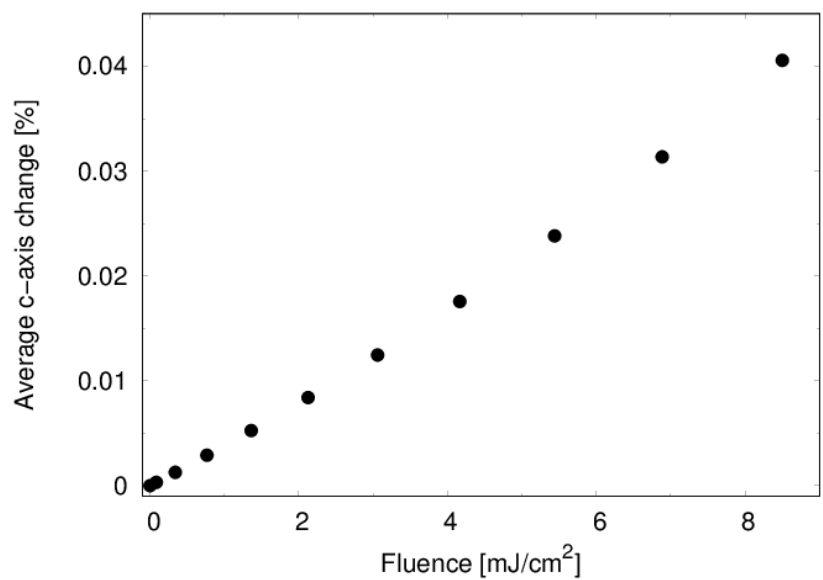

**Figure S4:** The approximately linear dependence of the average *c*-axis expansion as a function of fluence. The values on the *y*-axis are an averaging of the data from Fig. S3a (from 10 ps to 20 ps, where the expansion has reached a new equilibrium). Nonlinearities may indicate the contributions of other lattice anharmonicities, such as coupling to Raman phonons.

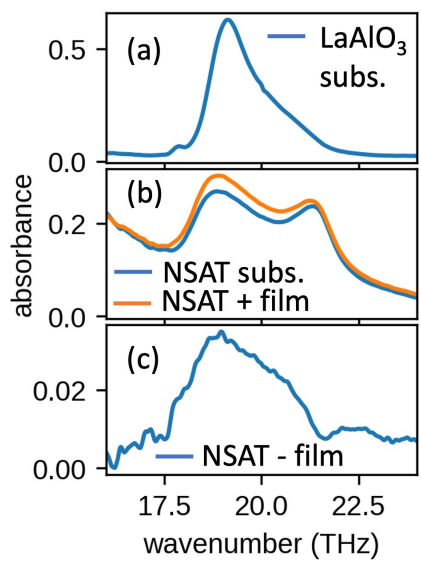

**Figure S5:** Fourier-transform infrared absorption data obtained from a bulk LaAlO$_3$ substrate, NSAT substrate, and the NSAT/LaAlO$_3$ sample system. (a) Shows the absorbance for LaAlO$_3$ bulk substrate. (b) Shows the absorbance of the sample including a LaAlO$_3$ thin film grown on NSAT and the bare NSAT substrate. (c) Shows the difference in absorbance between the NSAT/LaAlO$_3$ sample system and the bare NSAT substrate. The difference reports an absorbance of the thin LaAlO$_3$ film, when interference is neglected.

14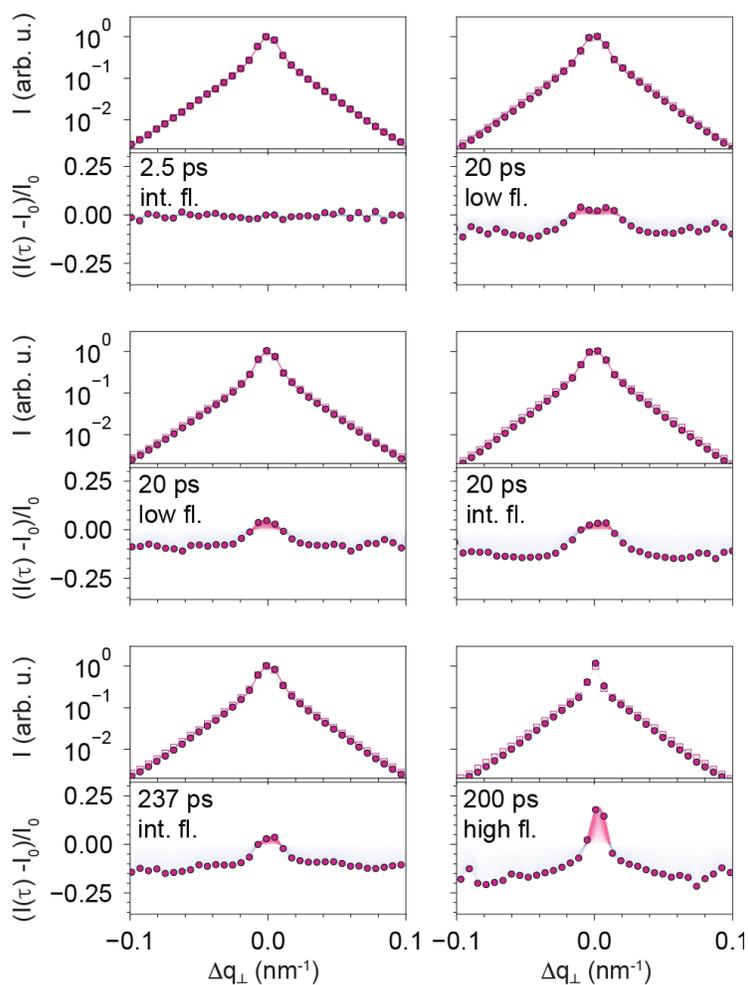

**Figure S6:** The change in the diffuse scattering as a function of time delay and fluence. No change in diffuse scattering is visible after 2.5 ps. At 20 ps, 200 ps, and 237 ps, the change in the diffuse scattering increases with the fluence.

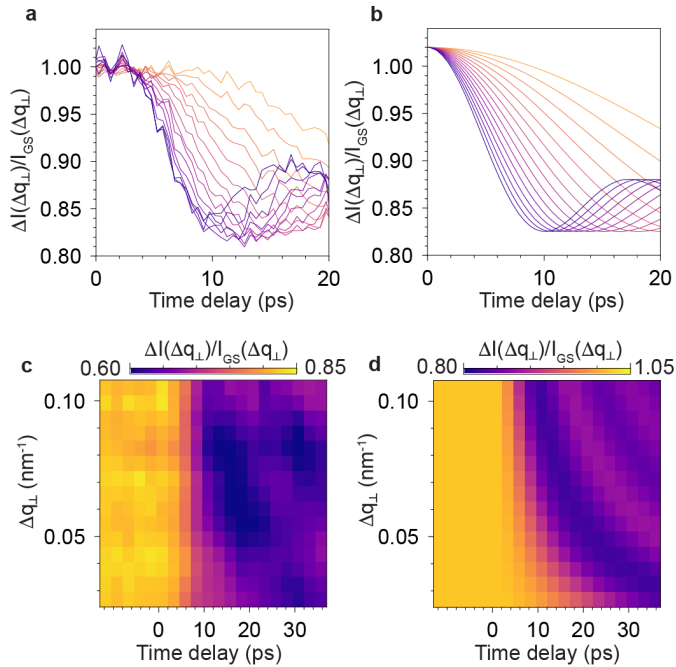

**Figure S7:** a) The data from Figure 5a of the main text, displayed line-by-line. (b) Fits to the data in (a), shown using the same color scheme. (c) Data similar to that in Figure 5a, but at a slightly later time delay, revealing a few oscillations (two oscillations are visible around 0.1 nm$^{-1}$). (d) The fit from (b) extended to longer time delays using the constant velocity assumption. In (c), the incident angle was offset from the Bragg peak. As the lattice expands, the peak shifts in θ, as described in the main text. Here, the shift reduces the intensity at a fixed θ. After approximately 5 ps, the shift stabilizes, allowing us to compare (d) and (c) after ~5 ps.

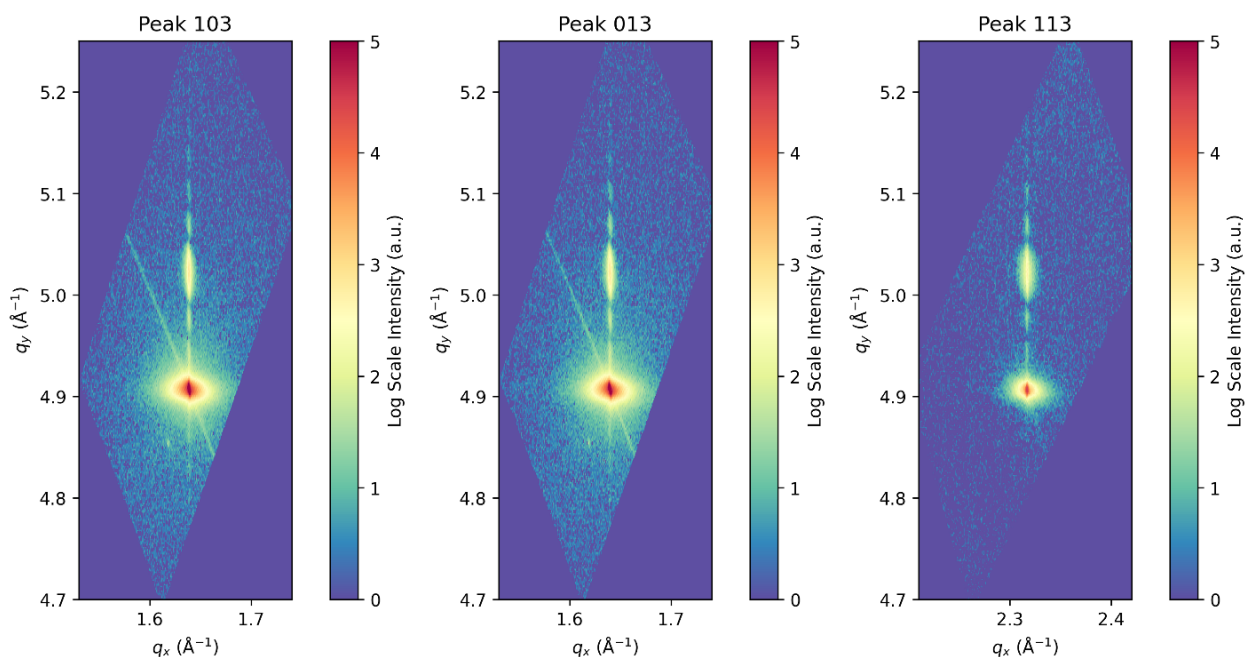

**Figure S8:** Reciprocal space maps around the (a) 103, (b) 013, and (c) 113 Bragg peaks showing the film is commensurately strained. This data was collected at ambient conditions using a PANalytical Empyrean diffractometer equipped with a Cu anode, PIXcel detector, a hybrid mirror-monochromator (2-bounce Ge channel-cut crystal), and ½ degree divergence slit. The miller indices are in pseudocubic perovskite notation.

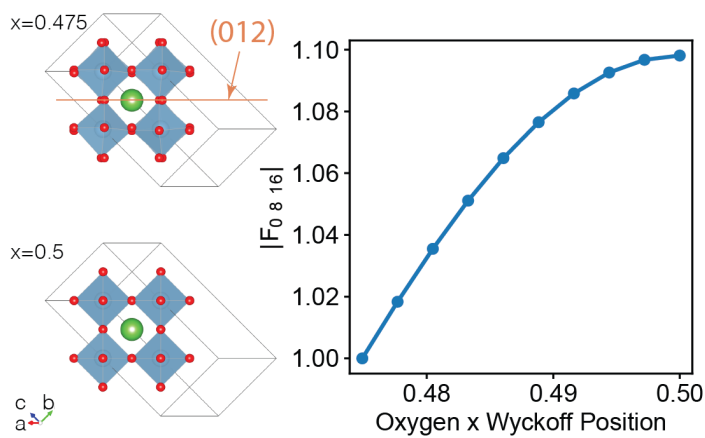

**Figure S9:** Magnitude of the structure factor corresponding to the cubic 008 peak measured in the experiment as a function of the octahedral tilt. We started with the rhombohedral bulk structure at room temperature[30] (top left) including octahedral rotations (retrieved from ICSD, [31]) and calculated the structure factor of the 008 Bragg peak (08L Bragg peak with L=16 in the rhombohedral space group, hexagonal setting). The octahedral distortions are controlled by the Wyckoff position of the oxygen atom along the x-axis; the transition to the cubic structure is associated with the x-position moving towards a value of 0.5. We modeled this transition by linearly varying x from its initial value of 0.475 to 0.5, showing that the structure factor changes by approximately 10% as the octahedra become unrotated. Assuming the order parameter in the Allen-Cahn model is 0.5−x, where x has two minima near 0.5, generates a plot consistent with the experimental data.